\documentclass[sigconf]{acmart}
\usepackage{tabularx}
\usepackage{enumitem}

\AtBeginDocument{%
  \providecommand\BibTeX{{%
    \normalfont B\kern-0.5em{\scshape i\kern-0.25em b}\kern-0.8em\TeX}}}

\copyrightyear{2020} 
\acmYear{2020} 
\setcopyright{rightsretained}
\acmConference[UMAP '20]{Proceedings of the 28th ACM Conference on User Modeling, Adaptation and Personalization}{July 14--17, 2020}{Genoa, Italy}
\acmBooktitle{Proceedings of the 28th ACM Conference on User Modeling, Adaptation and Personalization (UMAP '20), July 14--17, 2020, Genoa, Italy}
\acmPrice{15.00}
\acmDOI{10.1145/nnnnnnn.nnnnnnn}
\acmISBN{978-x-xxxx-xxxx-x/YY/MM}

\begin{document}

\title{Eliciting Touristic Profiles: A User Study on Picture Collections}

\author{Mete Sertkan}
\orcid{0000-0003-0984-5221}
\affiliation{%
  \institution{TU Wien}
  \streetaddress{Favoritenstrasse 9-11}
  \city{Vienna}
  \country{Austria}}
\email{mete.sertkan@tuwien.ac.at}

\author{Julia Neidhardt}
\orcid{0000-0001-7184-1841}
\affiliation{%
  \institution{TU Wien}
  \streetaddress{Favoritenstrasse 9-11}
  \city{Vienna}
  \country{Austria}}
\email{julia.neidhardt@tuwien.ac.at}

\author{Hannes Werthner}
\affiliation{%
  \institution{TU Wien}
  \streetaddress{Favoritenstrasse 9-11}
  \city{Vienna}
  \country{Austria}}
\email{hannes.werthner@tuwien.ac.at}

\renewcommand{\shortauthors}{Sertkan et al.}

\begin{abstract}
Eliciting the preferences and needs of tourists is challenging, since people often have difficulties to explicitly express them -- especially in the initial phase of travel planning.
Recommender systems employed at the early stage of planning can therefore be very beneficial to the general satisfaction of a user.
Previous studies have explored pictures as a tool of communication and as a way to implicitly deduce a traveller's preferences and needs.
In this paper, we conduct a user study to verify previous claims and conceptual work on the feasibility of modelling travel interests from a selection of a user's pictures. 
We utilize fine-tuned convolutional neural networks to compute a vector representation of a picture, where each dimension corresponds to a travel behavioural pattern from the traditional Seven-Factor model.
In our study, we followed strict privacy principles and did not save uploaded pictures after computing their vector representation.
We aggregate the representations of the pictures of a user into a single user representation, i.e., touristic profile, using different strategies.
In our user study with 81 participants, we let users adjust the predicted touristic profile and confirm the usefulness of our approach.
Our results show that given a collection of pictures the touristic profile of a user can be determined. 
\end{abstract}



\maketitle

\section{Introduction}
Nowadays, recommender systems (RS) accompany users in their everyday life, by proactively providing  ``suggestions for items a user may wish to utilize'' \cite{ricci2015recommender}.
Depending on the domain of application they face specific challenges.
In the tourism domain the recommended items are complex (i.e., they typically combine accommodation, transportation, activities, food, etc.), mostly intangible, and highly related to emotional experiences \cite{werthner1999information, Werthner:2004:ET:1035134.1035141}.
Thus, people have often difficulties to explicitly express their preferences, needs, and interests, especially in the initial phase of travel decision making \cite{zins2007exploring}. Recommender systems need to be transparent and explainable to the user in their selection process, otherwise they only increase the complexity of choosing a product.

In this study, we build on previous research to provide and evaluate an implicit picture-based preference elicitation method for a tourism-specific recommender system.
We employ computer vision models to determine the touristic profile of users, which are explainable to the users and where users are able to compare and adjust the results with their own preferences.
Our main contribution is a user study with 81 participants, who uploaded three to seven pictures and adjusted the automatically determined touristic profile and filled out a questionnaire about the usefulness of our system. 

Previous studies have shown that it is reasonable to follow the idiom ``a picture is worth a thousand words''. 
 
Neidhardt et al. \cite{Neidhardt:2014:EUU:2645710.2645767, neidhardt2015picture} used a simple picture selection process to determine the user's personality and travel preferences in an implicit gamified way.
In their approach, a user has just to select three to seven pictures out of a predefined fixed set of 63 pictures. Ferwerda et al. \cite{ferwerda2015predicting, ferwerda2018predicting} used visual features (e.g., brightness, saturation, etc.) and content features (i.e., identified concepts) of Instagram pictures to predict the personality of users.
Figuerdo et al. \cite{figueredo2018photos} classified users based on pictures of their social media streams into basic tourist classes.
However, previous research mainly focused on profiling the user, but in an ideal case picture-based approaches should be applicable on both users and items in order to characterize them in a match-able way. 

Addressing this issue, we introduced a more generic concept, where we utilize any kind of picture collections \cite{sertkan19pictures, sertkan2020pictures}.
Thus, depending on the source of the collection (e.g., social media stream, pictures provided by a destination management organization, etc.) either a user or an item (in this case a tourism destination) is characterized.
In this paper we deploy our conceptual work \cite{sertkan19pictures, sertkan2020pictures} with a user study to evaluate the user profile generation.
Additionally, we also consider the order of the pictures in a collection.

The main contributions of this paper are as follows:

\begin{itemize}
    \item We organize and evaluate a user study to determine the touristic profile of users based on pictures;
    \item We provide evidence that the touristic profile of users can be determined by a picture collection they provide;
    \item We analyze the difference between perceived profile and predicted profile;
    \item We compare the performance of our approach with and without considering the picture order.
\end{itemize}{}

The remainder of the paper is organised as follows:
In Section~\ref{sec:Background} we give an overview of the related work and focus on touristic preference models and picture-based approaches.
In Section~\ref{sec:Methods} we introduce our extension, illustrate the experimental setup and detail the evaluation.
In Section~\ref{sec:Results} we present the results and in Section~\ref{sec:Discussion} we discuss findings and implications.
Finally, in Section~\ref{sec:Conclusion} we conclude our work and provide some future outline.

\section{Background}
\label{sec:Background}

In this section we present related studies on user preference representation in the tourism domain and previous picture-based preference and/or personality elicitation approaches.

\subsection{Touristic Preference Models}
\label{sec:seven-factors}
Due to the complex nature of tourism products (e.g., a bundle of accommodation, transportation, activities, etc.), their strong ties to emotional experiences and high consumption costs, recommending the right product to a tourist is a non-trivial task.
Content about the tourism products and knowledge about the domain are crucial for RSs to bundle and recommend appropriate items \cite{neidhardt2015picture}.
Also, Burke and Ramezani \cite{Burke2011} suggest that content-based and / or knowledge-based paradigms are most appropriate for tourism recommenders.

For both paradigms capturing the preferences and needs of users is critical.
Especially, in case of the content-based recommendation paradigm defining an appropriate domain model and a distance measure in order to find matching products is essential.
Often users and items are characterized through a multidimensional vector space model, where each dimension covers a different touristic aspect.
In some cases, the vector space model is derived from the data and / or the structure of the data.
For instance, the vector space model in \cite{dietz2019designing} has the following dimensions: \textit{Arts \& Entertainment}, \textit{Food}, \textit{Nightlife}, \textit{Outdoors \& Recreation}, \textit{Venues}, \textit{Cost Index}, \textit{Temperature}, and \textit{Precipitation}.
Those dimensions are given / derived from different data sources they use (e.g., Foursquare, weather data, etc.).
In other cases, the vector space model is derived from the literature, like the Seven-Factor Model \cite{Neidhardt:2014:EUU:2645710.2645767, neidhardt2015picture}, which we use in this work.
The Seven-Factor Model combines the ``Big Five'' personality traits~\cite{goldberg1990alternative} (representing the long-term preferences) and 17 tourist roles of Gibson and Yiannakis \cite{GIBSON2002358} (representing the short-term preferences).
Seven basic factors were obtained by reducing the initial 22 (i.e., 5 + 17) dimensions via factor analysis.
Thus, the factors of the Seven-Factor model are considered as independent from each other.
Furthermore, users are depicted as a mixture of the Seven-Factors (since people can have different tastes) rather than classified to only one factor.
For a better understanding, the Seven-Factors can be briefly summarized as follows \cite{Neidhardt:2014:EUU:2645710.2645767, neidhardt2015picture}:

\setlist[description]{font=\normalfont\itshape\space}

\begin{description}
\item[Sun \& Chill-Out (F1) -]
	a neurotic sun lover, who likes warm weather and sun bathing and does not like cold, rainy or crowded places;
\item[Knowledge \& Travel (F2) -]
	an open minded, educational and well-organized mass tourist, who likes travelling in groups and gaining knowledge, rather than being lazy;
\item[Independence \& History (F3) -]
	an independent mass tourist, who is searching for the meaning of life, is interested in history and tradition, and likes to travel independently, rather than organized tours and travels;
\item[Culture \& Indulgence (F4) -]
	an extroverted, culture and history loving high-class tourist, who is also a connoisseur of good food and wine;
\item[Social \& Sports (F5) -]
	an open minded sportive traveller, who loves to socialize with locals and does not like areas of intense tourism;
\item[Action \& Fun (F6) -]
	a jet setting thrill seeker, who loves action, party, and exclusiveness and avoids quiet and peaceful places;
\item[Nature \& Recreation (F7) -]
	a nature and silence lover, who wants to escape from everyday life and avoids crowded places and large cities.
\end{description}

\subsection{Picture-Based Approaches}
\label{sec:picture-related}
Historical user data and knowledge about the user's preferences and needs are essential for RSs in order to provide good recommendations.
Often, this information is not available, for example, if the user is logged out or generally for any new user.
This issue is also known as the ``cold start'' problem.
In this case one can elicit the preferences and needs of users explicitly (e.g., by asking related questions, etc.) or implicitly (e.g., by observing the behaviour, etc.)~\cite{sertkan_what_2019}.
However, people often have difficulties in explicitly expressing their travel preferences, which is due to the complexity of the tourism products.
Thus, implicit preference elicitation techniques are promising in such cases. 

Previous research demonstrated that it is reasonable to use pictures as a medium for communication between a user and a recommendation system.
In this way the user is addressed on an emotional implicit level and thus an explicit preference statement is not needed \cite{neidhardt2015picture}. 

Ferwerda et al. \cite{ferwerda2015predicting, ferwerda2018predicting} showed that the personality of people can be determined through their Instagram pictures.
In \cite{ferwerda2015predicting} they only used low level features, such as brightness and saturation, to predict the well-known ``Big Five'' personality traits, i.e., \textit{openness}, \textit{conscientiousness}, \textit{extraversion}, \textit{agreeableness}, \textit{neuroticism}.
They also utilized high level features, i.e., concepts they identified through Google Vision API\footnote{\url{https://cloud.google.com/vision}} \cite{ferwerda2018predicting}.
Furthermore, personality traits tend to be stable over time and thus can facilitate the prediction of the long-term behaviour of people \cite{matthews_deary_whiteman_2003,WOSZCZYNSKI2002369}.
Since the application domain of our work is tourism, personality traits only are not sufficient to make recommendations.
Therefore, we use the Seven-Factor Model, which is covering personality and touristic traits. 

Our work can be seen as a continuation of the picture-based approach of Neidhardt et al. \cite{Neidhardt:2014:EUU:2645710.2645767, neidhardt2015picture}.
In their approach they use a simple gamified picture selection process to depict users within the Seven-Factor Model.
In the mentioned picture selection process people have to select three to seven pictures out of a given set of 63 pre-defined pictures and based on their selection the Seven-Factors are calculated.
The set of 63 pictures and the loading of each picture with the Seven-Factors were identified through workshops and experts.
The fundamental difference to our approach is that we do not limit the user interaction to a fixed set of pictures.

Figuerdo et al. \cite{figueredo2018photos} use convolutional neural networks (CNNs) to characterize people based on pictures.
They utilize pictures from people's social media stream (i.e., Facebook, Instagram, and Google Plus) to classify them into five basic classes, i.e., \textit{Historical/Cultural}, \textit{Adventure}, \textit{Urban}, \textit{Shopping}, and \textit{Landscape}.
Similar to the approach of Ferwerda et al. \cite{ferwerda2018predicting} they identify concepts (i.e., scenes) in the pictures and based on the frequency of these concepts they use a fuzzy classifier to assign scores to the classes.
They use CNNs in order to identify the concepts and in turn to determine scores for their classes, whereas in our approach we train CNNs to directly output the Seven-Factor scores. Our approach in a single model prevents information loss of the intermediate step.
Furthermore, we utilize a touristic preference model derived from the literature, whereas the classes in \cite{ferwerda2018predicting} seem to be arbitrarily defined.
Finally, in our approach we let the user decide, which pictures should be used for their travel profile generation. 

Previous studies mainly concentrate on the one-way profiling of pictures to user models. In contrast, our goal is to employ a generic profiler, which we already conceptually introduced in \cite{sertkan19pictures, sertkan2020pictures} and which can universally characterize users and recommendation items based on corresponding picture collections in a comparable way.

\section{Methods}
\label{sec:Methods}

In this section we describe the ``generic profiler'' \cite{sertkan19pictures, sertkan2020pictures} plus the extension for also accounting the order of pictures in a collection.
Furthermore, we present the experimental setup of the user study and finally, we define our evaluation procedure and metrics.

\subsection{A Generic Profiler}

\begin{figure}[h]
  \centering
  \includegraphics[width=\linewidth]{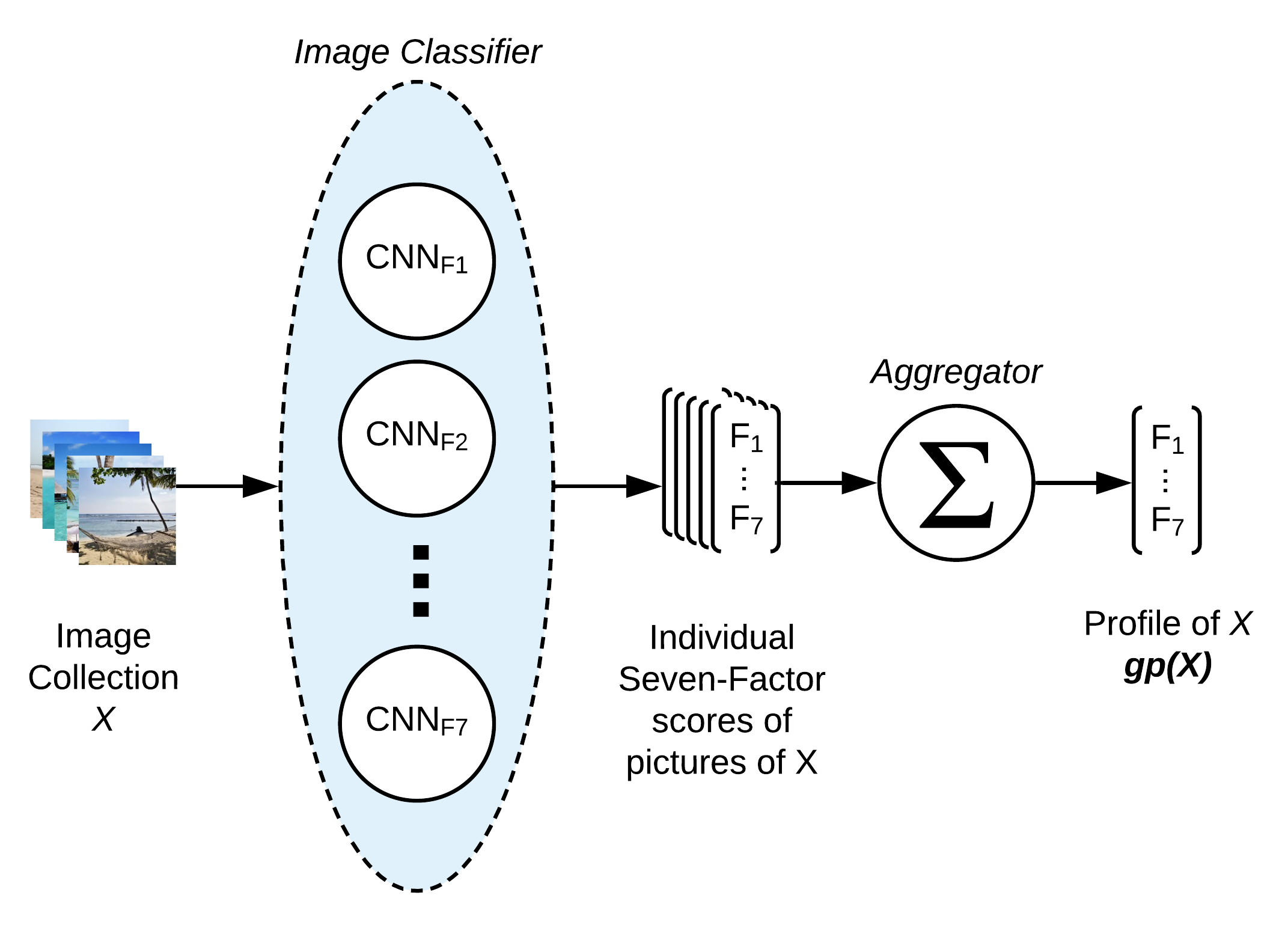}
  \caption{Eliciting the Seven-Factors from picture collections - A generic approach \cite{sertkan19pictures,sertkan2020pictures}.}
\end{figure}

Given a collection of pictures as input the generic profiler determines the collection's Seven-Factor representation, i.e., the touristic profile.
This is realized in two main steps, namely \textit{Classification} and \textit{Aggregation}. 
The purpose of the Classification step is to determine the Seven-Factor representation of an input picture.
Each factor in the Seven-Factor Model is treated independently and therefore seven CNNs are trained as binary classifier.
The output of each classifier (i.e., class probability) is used as the score of the corresponding factor.
All seven scores are then combined into one seven-dimensional vector, i.e., the Seven-Factor representation of the input picture.
For each binary classifier a pretrained ResNet50 model \cite{He_2016_CVPR} is adapted and fine-tuned \cite{sertkan19pictures,sertkan2020pictures}.

Given a collection of pictures as input the Classification step returns the Seven-Factor representation $f^p$ for each picture in the input collection.
Therefore, the main role of the Aggregation step is to aggregate the individual Seven-Factor representations $f^p_i$ of a collection $X$ with $N$ pictures into one representation, which characterizes the whole collection \cite{sertkan19pictures,sertkan2020pictures}.
In \cite{sertkan19pictures,sertkan2020pictures} the aggregation is proposed as a simple mean.
Therefore, the ``generic profile'' $gp(X)$ of a collection $X$ is defined as following:

\begin{equation}
  \label{eq:AVG}
  gp(X)=\frac{1}{N}\sum_{i=1}^{N}f^p_i
\end{equation}

In addition to the simple mean we also consider in this work the order of the pictures within a collection for the aggregation (more details in Section~\ref{sec:WeightedAverage}) and then compare both aggregation strategies.

\subsection{Accounting for the Picture Order}
\label{sec:WeightedAverage}
Insights of a study conducted in \cite{neidhardt2015picture} show that most people tend to select three to seven pictures out of a given set of pictures.
Furthermore, the order of the pictures might carry valuable information, since in the same study people often re-ranked their initial selection.
In order to consider the order (i.e., rank) of the pictures in the user's selection they experimented with different strategies.
The best strategy not only considered the rank of the pictures in the user's selection, but also the number of pictures in the user's selection.

We adapt those insights and also follow the best performing strategy by 1) Limiting the size of the input collection to minimum three and maximum seven pictures; and 2) Aggregating the individual Seven-Factor scores $f^p_i$ (i.e., output of the \textit{Classification} step) of an input collection $X$ with $n=3, ..., 7$ pictures through weighted averaging. Thus, the profile of $X$, i.e., $gp(X)$, is defined as follows:

\begin{equation}
  \label{eq:WA}
  gp(X)=\frac{\sum_{i=1}^{n}\omega_if^p_i}{\sum_{i=1}^{n}\omega_i}
\end{equation}
\begin{equation}
  \label{eq:W}
  \omega_i=7\frac{-r + n + 1}{\sum_{k=1}^{n}k}
\end{equation}

where $\omega_i$ is the weight of each picture and depends on the collection size $n$ and the rank $r$ of the considered picture. For instance,  $\omega_i$ for the first ranked picture in a collection of three pictures equals to $\frac{21}{6}$, $\frac{14}{6}$ for the second ranked picture, and finally $\frac{7}{6}$ for the third ranked picture.
The sum of all weights always equals seven.

\subsection{Experimental Design}

\begin{figure}[h]
  \centering
  \includegraphics[width=\linewidth]{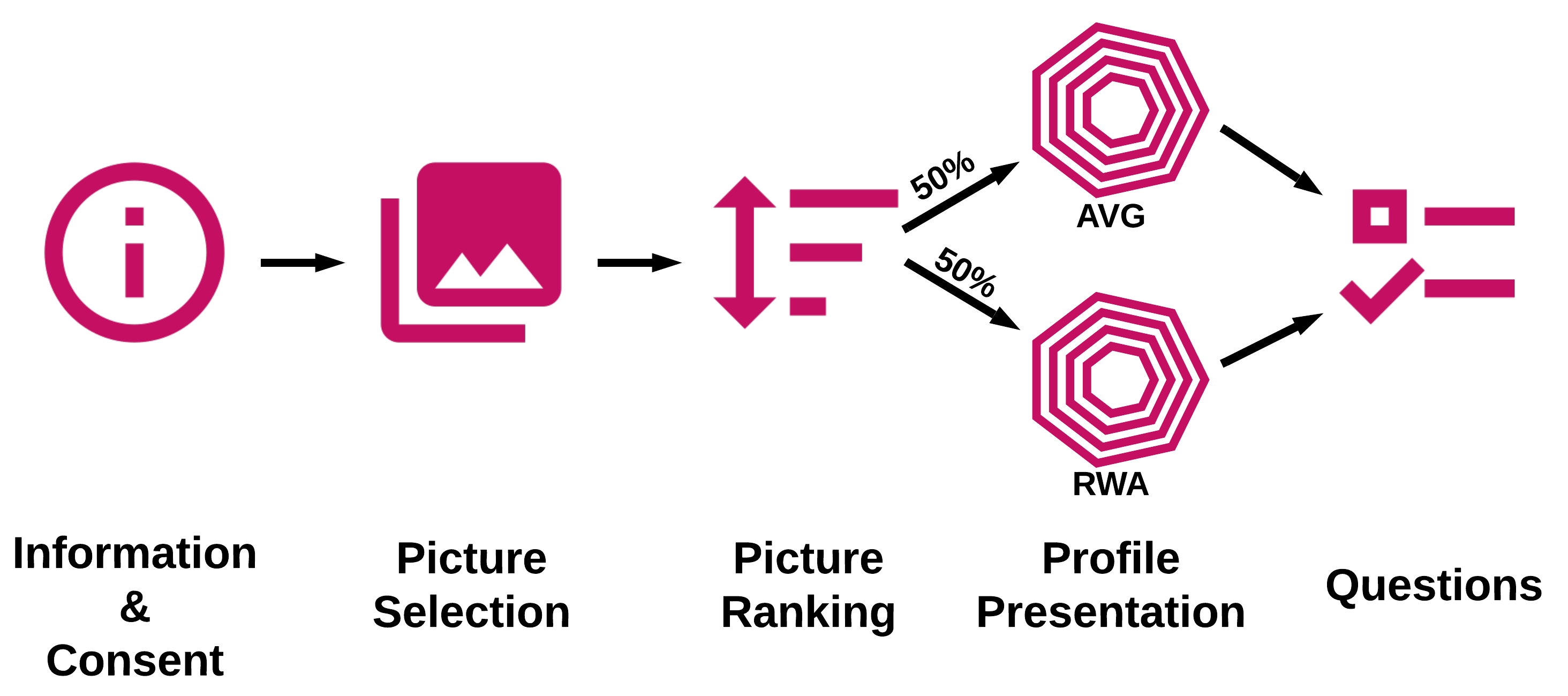}
  \caption{Study Procedure - Consisting of following steps: 1) Information and Consent; 2) Picture Selection; 3) Picture Ranking; 4) Profile Presentation; 5) Questions.}
  \label{fig:StudyProcedure}
\end{figure}

Figure~\ref{fig:StudyProcedure} illustrates the main steps of the study procedure.
Participants start at the landing page, i.e., Step 1, where we present detailed information about the study and a consent form for participation.
In Step 2, we ask the participants to imagine their next hypothetical vacation and based on this to select three to seven pictures, i.e., either their own pictures or pictures downloaded from the web.
We clearly state that the pictures used in this study are not saved or displayed at any time.
In Step 3, we ask the participants to rank the selected pictures according to their relevance.
Based on the selected pictures we present the resulting touristic profile, i.e., Seven-Factor representation, to the participant in Step 4.
The shown profile is, with equal chance, either based on the simple average aggregation strategy hereinafter referred to as \textit{AVG} (see Equation~\ref{eq:AVG})  or based on the rank weighted averaging aggregation strategy hereinafter referred to as \textit{RWA} (see Equations~\ref{eq:WA} and \ref{eq:W}).
We also provide a brief explanation of the Seven-Factors plus a link to a more detailed description of the Seven-Factor Model.
Finally, in Step 5 we ask the participants following questions (*~marks mandatory questions):

\begin{description}
    \item[Q01 -]* It was easy to find 3 to 7 pictures.
    \item[Q02 -]* I mainly used pictures downloaded from the internet (e.g., Google, Flickr, etc.).
    \item[Q03 -]* I mainly used my own pictures.
    \item[Q04 -]* I understood the explanations of the Seven-Factors.
    \item[Q05 -]* The resulting profile matches my preferences.
    \item[Q06 -]  Which factor in the resulting profile does not match well? (multiple answers allowed)
    \item[Q07 -] How would you adjust the resulting profile? (multiple adjustments allowed)
    \item[Q08 -]* What is your age?
    \item[Q09 -]* What is your gender?
    \item[Q10 -]* What is your highest degree of education?
    \item[Q11 -]* How often do you travel for pleasure (leisure/tourism)?
    \item[Q12 -]  Comments/Suggestions.
\end{description}

For questions \textit{Q01}-\textit{Q05} we provide a five-point Likert scale ranging from \textit{strongly disagree} to \textit{strongly agree}.
For question \textit{Q06} we provide seven checkboxes, each for one factor of the Seven-Factor Model.
In case of \textit{Q07}, we provide seven sliders (again each for one factor of the Seven-Factor Model), where the values are pre-set to the Seven-Factor scores of the predicted touristic profile.
Questions \textit{Q08}-\textit{Q11} can be answered via Radio buttons, where we always provide the option ``prefer not to say''.
Question \textit{Q12} is an open question, which can be answered via text field. 

The questions can be related to three main topics:
1) Picture selection \textit{Q01-Q03}, where we focus on picture source and difficulty to find pictures; 2) The touristic profile \textit{Q04-Q07}, where we concentrate on the overall performance plus capturing the difference between perceived and predicted characteristics; and finally, 3) Demographics \textit{Q08-Q011}.

Besides the explicitly asked questions, we also track following: 
\begin{itemize}
    \item Time spend for the picture selection and ranking process;
    \item Time spend for understanding the profile and answering the questions;
    \item Number of picture re-rankings. 
\end{itemize}

\subsection{Evaluation}
The purpose of questions \textit{Q01}-\textit{Q04} and \textit{Q08}-\textit{Q11} is to get more insights about the participants and their behaviour, and to find support for generalizability statements.
Also, tracking interactions and time might give insights about the behaviour and hints about difficulties participants face.
Altogether, those insight can be used to further improve the introduced concept and moreover its presentation (i.e., user interface).

Questions \textit{Q05}-\textit{Q07} are used to assess the overall performance and also the difference in performance with respect to the aggregation strategies (i.e., \textit{AVG} and \textit{RWA}).
We use the mean absolute error (\textit{MAE}) in order to assess the difference in each factor of the Seven-Factor Model between predicted touristic profile and the user's perception (i.e., user's adjustment to the presented profile).
Besides considering the predictive performance in each factor, we also treat the user's touristic profile as such by considering its distance to the  user's perception. 
Therefore, we use Kendall's Tau distance ($DIST_{\tau}$), Spearman's Footrule ($DIST_{SPEAR}$), and the Euclidean distance ($DIST_{EUCL}$). 

\begin{figure}[h]
  \centering
  \includegraphics[width=.7\linewidth]{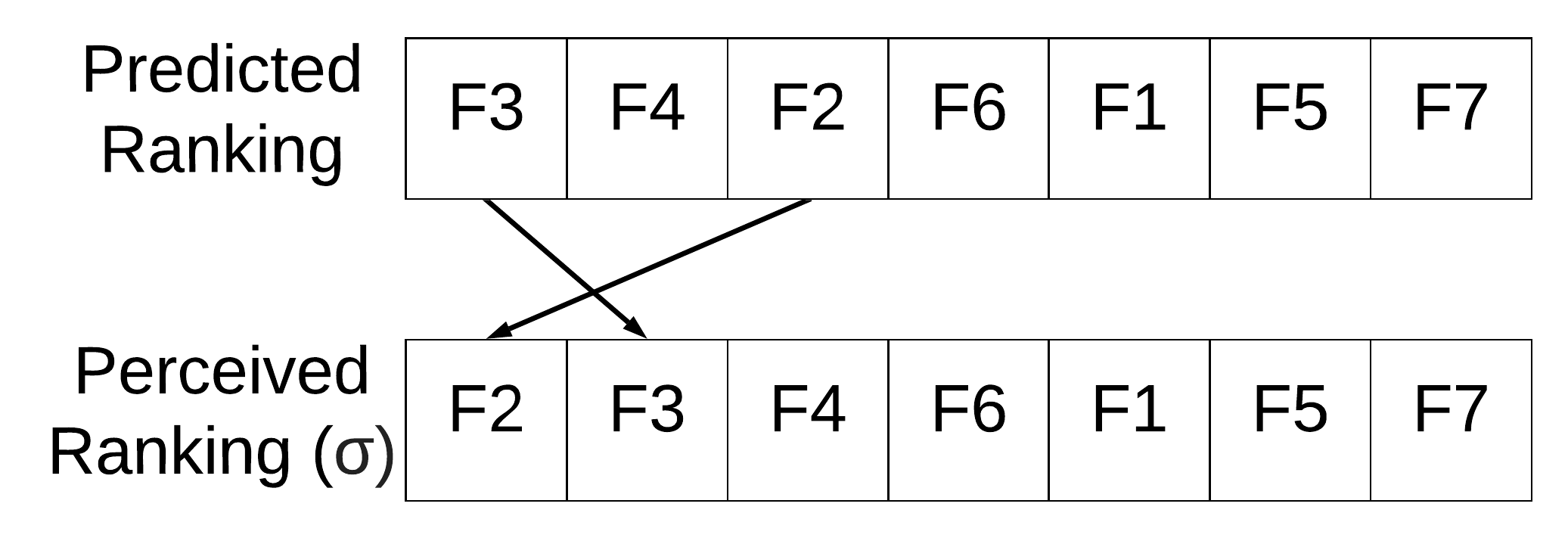}
  \caption{Kendall's Tau distance - Total number of inversion in $\sigma$. Note, in this example Kendall's Tau distance is 2. }
  \label{fig:KendallsTau}
\end{figure}

Comparing the predicted ranking (i.e, rank of the factors based on scores in the predicted profile) and perceived ranking ($\sigma$) (i.e., rank of the factors based on scores in the adjusted profile), the Kendall's Tau distance can be interpreted as the total number of inversions in $\sigma$ (see Figure~\ref{fig:KendallsTau}) \cite{dwork2001rank}.
Here, a pair of elements $F_i$ and $F_j$ is considered as inversed if $R_{F_i} > R_{F_j}$ and $R_{\sigma(F_i)}<R_{\sigma(F_j)}$, where $R_{F_i}$ stands for the ranking of an element and $R_{\sigma(F_i)}$ the ranking of an element in $\sigma$.
Thus the Kendall's Tau distance is defined as following: 

\begin{equation}
  \label{eq:Tau}
  DIST_{\tau}=\sum_{R_{F_i}<R_{F_j}}1_{R_{\sigma(F_i)}>R_{\sigma(F_j)}}
\end{equation}

\begin{figure}[h]
  \centering
  \includegraphics[width=.7\linewidth]{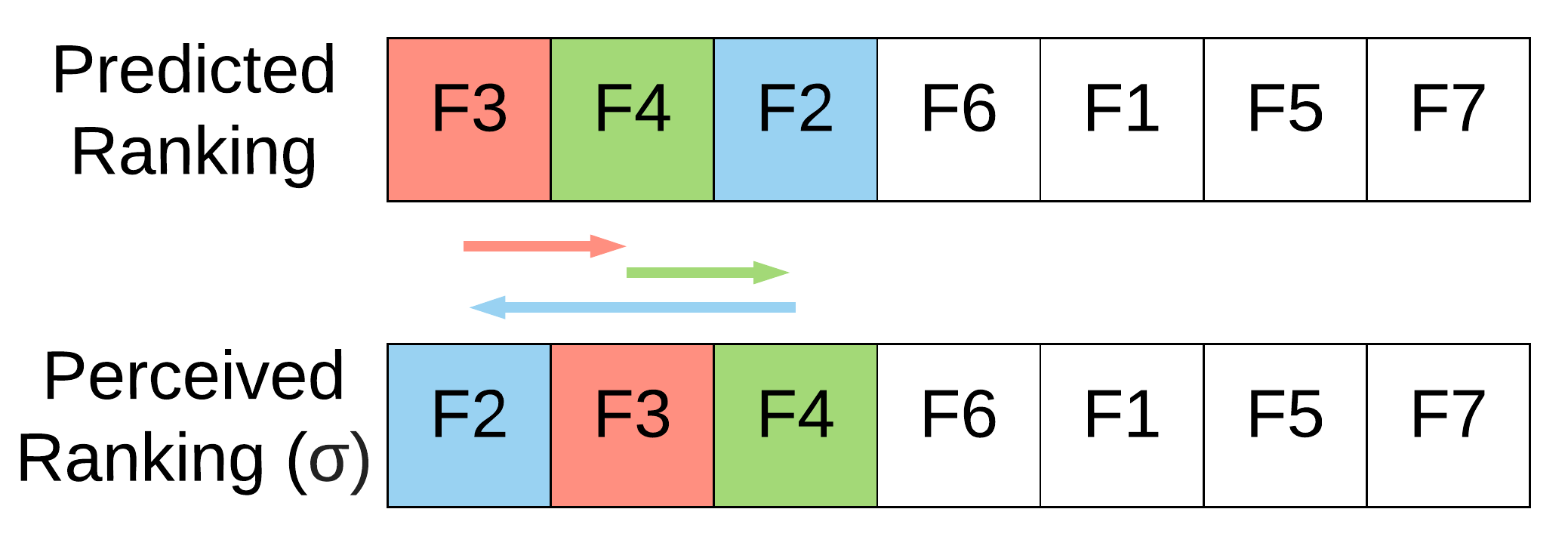}
  \caption{Spearmans's Footrule distance - Total displacements of elements in $\sigma$. Note, in this example Spearman's Footrule distance is 4. }
  \label{fig:SpearmansFootrule}
\end{figure}

On the other hand, the Spearman's Footrule distance (see Figure~\ref{fig:SpearmansFootrule}) can be interpreted as the total number of displacement of all elements \cite{dwork2001rank}.
Here, a displacement is considered as the distance an element $F_i$ has to be moved to match $\sigma(F_i)$, which can also be written as $|R_{F_i} - R_{\sigma(F_i)}|$.
Since the Spearman's Footrule is defined as the total number of displacements, it can be written as following:

\begin{equation}
  \label{eq:Spear}
  DIST_{SPEAR}=\sum_{i}|R_{F_i} - R_{\sigma(F_i)}|
\end{equation}

By comparing the ranking, we account for the change in relevance of each factor of the Seven-Factor Model, for instance, the factor \textit{Sun \& Chill-Out} might be more relevant (i.e., ranked higher) in the user's perception than in the predicted touristic profile.
Besides that, we also consider the distance of the presented and the perceived touristic profile based on the actual difference in Seven-Factor scores by using the Euclidian distance, which is defined as following:

\begin{equation}
  \label{eq:EUCL}
  DIST_{EUCL}=\sqrt{\sum_{i=1}^{7}(predicted\_F_i - perceived\_F_i)^2}
\end{equation}

Finally, in order to identify significant distributional differences between the predicted Seven-Factor scores and the perceived Seven-Factor scores, we use the paired Student's t-test or the Wilcoxon signed-rank test depending on the outcome of the Shapiro-Wilk normality test.
Furthermore, to compare differences based on the two aggregation strategies (i.e., \textit{AVG} and \textit{RWA}) and the nature of the considered variable we either use Mann-Whitney U test or Fisher's exact test.

\section{Results}
\label{sec:Results}

In this section we present and analyse the outcomes of the conducted user study.
We provide insights about the people who participated in the study.
We analyse the picture selection and ranking process.
Finally, we evaluate the performance of our model and compare differences to the user's perception and differences in outcome of both aggregation strategies.

\subsection{Participants}
\label{sec:participants}
The participants were recruited in January 2020 by i) sharing the user study at the ENTER2020 international eTourism conference and through international mailing lists of tourism experts, and ii)~on social media, and among friends and colleagues, with no substantial differences in the results.
In total, 81 participants finished the user study, where 62\% of the participants defined them self as man and 38\% as woman.
Their self-reported age distribution looks like following: 60\% 25-34 years; 20\% 35-44 years; 7\% 45-55 years; 7\% above 55 years; 3\% 18-24 years; and 3\% below 18 years.
The vast majority of the participants (i.e., 87\%) reported that the highest educational degree they hold is  either bachelor, master, or PhD degree and furthermore, 9\% answered with high school degree, 2\% with less than a high school degree, and 2\% with ``other''. 
The majority of participants (i.e., 62\%) reported that they travel between one and three times a year (i.e., they chose the option 1-2 times a year or 2-3 times a year) for pleasure (tourism/leisure), 15\% answered with 3-4 times a year, 4\% with 4-5 times a year, 12\% with more than 5 times a year, and finally 7\% with less than one time in a year. 
About 42\% of the participants used a mobile device.
Finally, 90\% of the participants reported that they understood the description of the Seven-Factors (i.e., agreement or strong agreement with \textit{Q04}).

As already mentioned, the touristic profile (i.e., Seven-Factor representation) shown to the user is either based on the \textit{AVG} aggregation strategy or on the \textit{RWA} aggregation strategy, with equal chance and randomly assigned.
From 81 participants in total, 51\% (N=41) were assigned to \textit{AVG} and 49\% (N=40) to \textit{RWA}.
Note, in the following sections we discuss the outcomes with respect to all participants (N=81) and broken down by both \textit{AVG} (N=41) and \textit{RWA} (N=40).

\subsection{Picture Selection \& Ranking}
\label{sec:picture-selection-ranking}
As already mentioned, we gave the participants the option so select between three and seven pictures.
The majority of the participants (i.e., 52\%) selected only three pictures, 16\% selected four pictures, 11\% six pictures, another 11\% seven pictures, and finally 10\% five pictures.
We also asked the participants to re-consider the initial ranking of the selected pictures, where only 21\% did actually a re-ranking.
Those, who considered a re-ranking, changed the initial ranking between one and four times.
Half of the participants finished the selection and ranking task within 2.7 minutes, 75\% of the participants completed after 5.8 minutes, and after 10.7 minutes already 90\% were finished.
The majority of the participants (72\%) agreed or strongly agreed with Q01 (i.e., ``It was easy to find 3 to 7 pictures''), which is in line with the reported timing above.

\begin{figure*}[h]
  \centering
  \includegraphics[width=\textwidth]{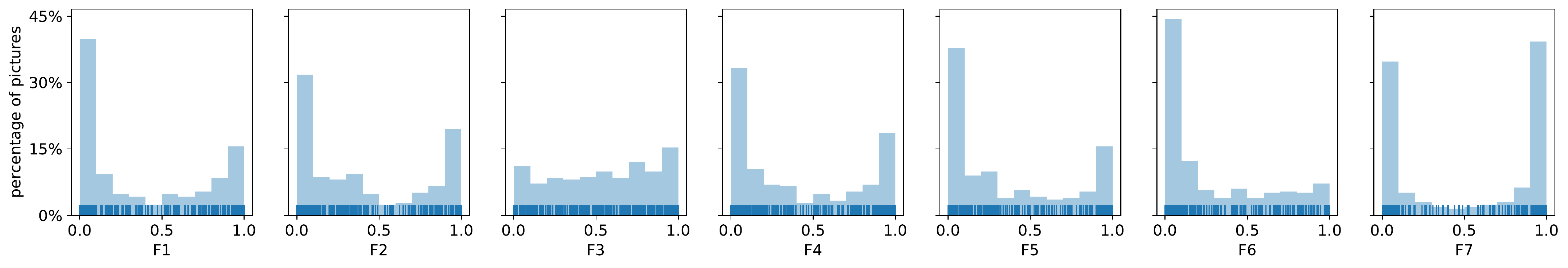}
  \caption{Distribution of Seven-Factor scores of the uploaded pictures.}
  \label{fig:UserPictureScoreDist}
\end{figure*}

The distributions of Seven-Factor scores of the uploaded pictures (see Figure~\ref{fig:UserPictureScoreDist}) have overall a similar shape, where either there are strong signals in the pictures for the considered factor (i.e., high score) or there are no signals (i.e., low score).
Except in case of factor \textit{Independence \& History (F3)}, where the scores are more evenly distributed.
Furthermore, there only were very few signals for the factor \textit{Action \& Fun (F6)} in the user provided pictures.
Both observations might indicate that some factors are harder to capture, leaving the room for further improvements.

In contrast to analysing the distribution of the Seven-Factor scores of all uploaded pictures, where we treated the Seven-Factors individually and all pictures at once, we also analysed the diversity of the provided pictures per user selection.
In other words, we investigated whether the uploaded images are homogeneous (e.g., only images of nature) or more diverse (e.g., images of nature, sports, and beach).
Here, diversity of a users picture selection is defined as the average of the pairwise distances of the pictures in the respective selection. We used $DIST_{\tau}$, $DIST_{SPEAR}$, and $DIST_{EUCL}$ as distance measure.

The results are listed in Table~\ref{tab:UsersPictureDiversity}, for instance, in case of the selection size of three one has to swap on average 8 times adjacent (based on ranking) factors in order to match the ranking of factors with another picture's ranking in the same selection.
Similarly, one has to move the factors (based on ranking) 13 times in order to match the ranking of factors with another picture's ranking in the same selection.
Also, the point-wise difference, i.e., diversity based on $DIST_{EUCL}$, is relatively high.
Thus, in case of picture selection size three, the participants selected relatively diverse pictures, which is also true in case of all other picture sizes.
However, the diversity in pictures in selections of sizes six or seven is relatively lower compared to diversity of pictures in selections of all other sizes, which is not expected (since there are more pictures to compare with). 

\begin{table}[h]
  \caption{
  Diversity in users' picture selection of different sizes.
  Note, ``\#Pics'' is selection size;  
  ``Kendall's'' is the mean of the average pairwise $DIST_{\tau}$ of the users' picture selection; ``Spearman's'' is the mean of the average pairwise $DIST_{SPEAR}$; and ``Euclidean'' is the mean of the average pairwise $DIST_{EUCL}$.} 
  \label{tab:UsersPictureDiversity}
  \begin{tabular}{lrrr}
    \toprule
    \#Pics & Kendall's & Spearman's &  Euclidean \\
    \midrule
    3 &   8.41 &  13.08 &    1.11 \\
    4 &   8.07 &  12.94 &    1.06 \\
    5 &  10.82 &  16.34 &    1.35 \\
    6 &   7.73 &  12.77 &    \textbf{1.00} \\
    7 &   \textbf{7.30} &  \textbf{11.47} &    1.03 \\
    \bottomrule
  \end{tabular}
\end{table}

\subsection{Overall Performance}
\label{sec:performance}

In order to assess the overall performance and thus user satisfaction, we asked the participants whether or not the presented touristic profile matched their preferences (\textit{Q05}) and which of the factors in the shown touristic profile did not match well (Q06).

Our approach got quite positive feedback, where 65\% of the participants were overall satisfied with the resulting touristic profile (i.e., agreement or strong agreement with \textit{Q05}). 
The distribution is also reflected in both strategies.
Note, the touristic profile presented to the user is either based on \textit{AVG} aggregation (N=41) or on \textit{RWA} aggregation (N=40) (the strategies were randomly assigned).
Table~\ref{tab:OverallSatisfaction} lists the summary statistics for the level of agreement to \textit{Q05} for both strategies and overall.
No significant difference could be shown between both strategies with respect to \textit{Q05}. 

\begin{table}[h]
  \caption{Summary statistics of level of agreement to Q05 - Overall and broken down by aggregation strategy. Note, 0 is strongly disagree and 4 is strongly agree.} 
  \label{tab:OverallSatisfaction}
    \begin{tabular}{lccccc}
    \toprule
    {} & mean & sd & min &  median & max \\
    \midrule
    Overall (N=81) & 2.69 & 0.81 & 0 & 3 & 4 \\
    AVG (N=41) & 2.68 & 0.81 & 0 & 3 & 4 \\
    RWA (N=40) & 2.70 & 0.81 & 1 & 3 & 4 \\
    \bottomrule
    \end{tabular}
\end{table}

\begin{figure}[h]
  \centering
  \includegraphics[width=\linewidth]{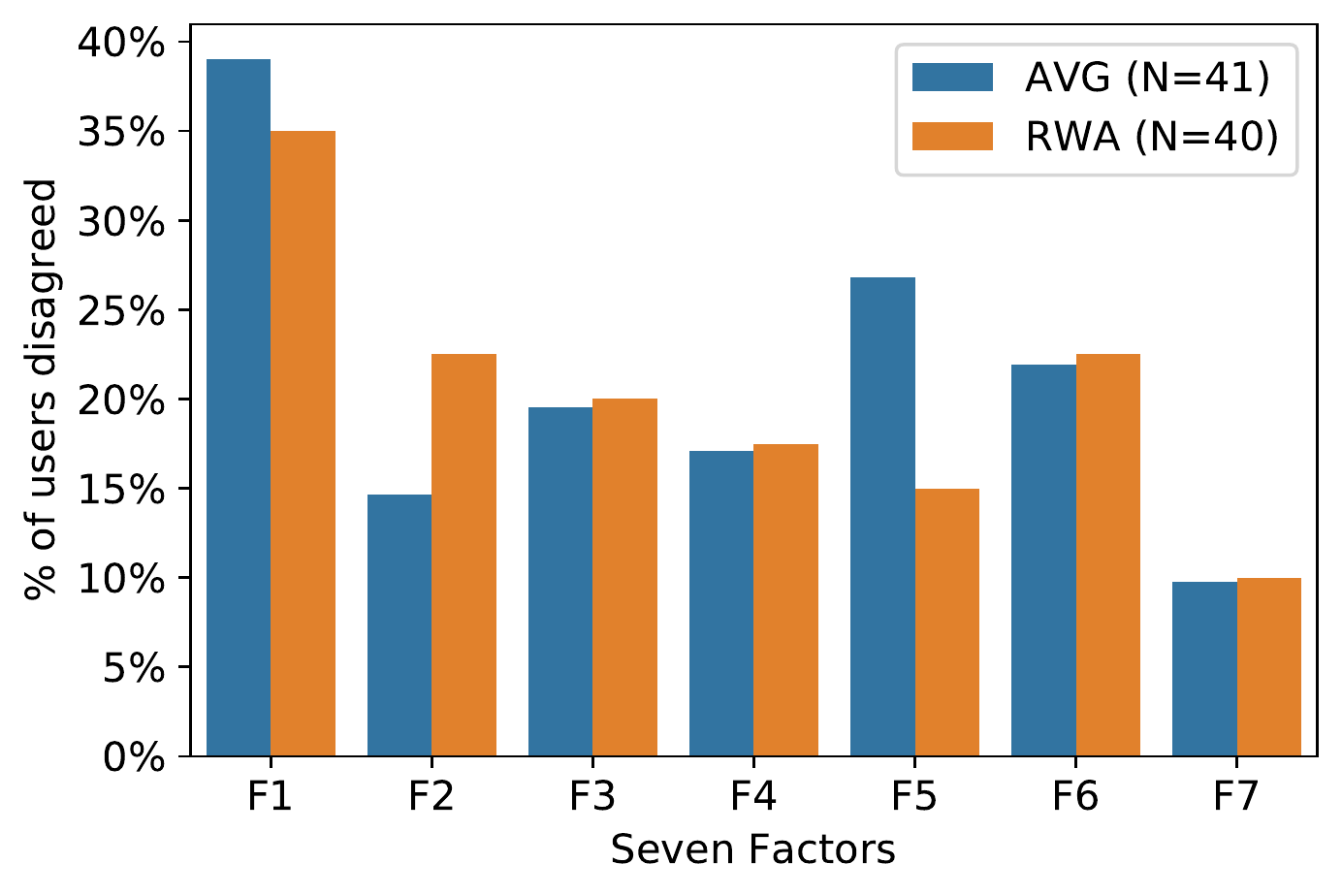}
  \caption{Disagreement with factors of the presented touristic profile (\textit{Q06}) broken down by aggregation strategy.}
  \label{fig:FactorDisagreement}
\end{figure}

The participants disagreed (i.e., checked the option ``did not match well'') the most with factor \textit{Sun \& Chill-Out} (37\% of the participants) and the least with factor \textit{Nature \& Indulgence} (10\% of the participants).
In all other factors of the Seven-Factor Model 17-22\% of the participants disagreed.
This also holds if the participants response to \textit{Q06} is viewed separately with respect to the both aggregation strategies, i.e., \textit{AVG} and \textit{RWA} (see Figure~\ref{fig:FactorDisagreement}). No significant difference could be shown between both strategies with respect to \textit{Q06}. 

\begin{figure*}[h]
  \centering
  \includegraphics[width=\textwidth]{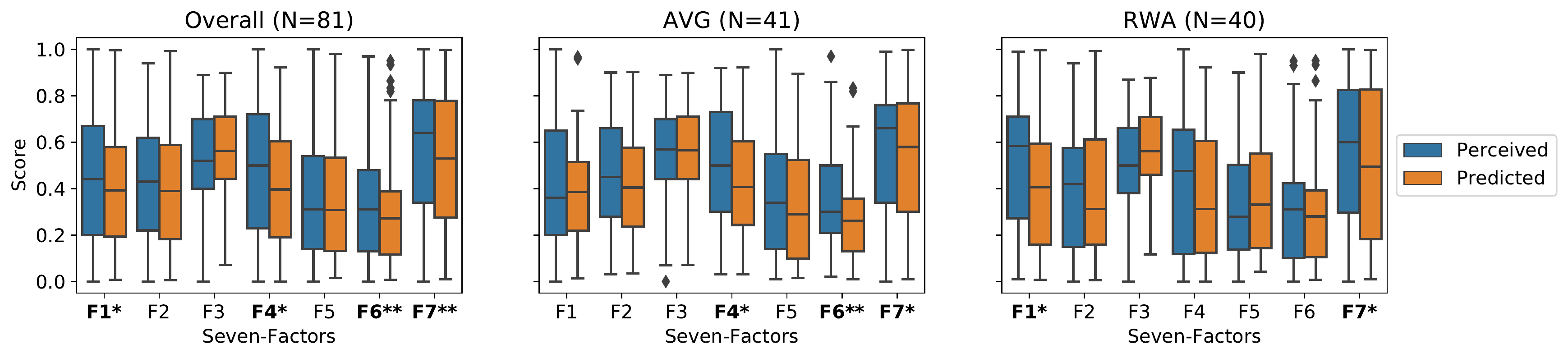}
  \caption{Predicted vs. Perceived Seven-Factor scores - Overall and broken down by aggregation strategy. Note, significant differences are highlighted (significance levels: * p<0.05 and ** p<0.01).}
  \label{fig:PredictedVsPerceivedDist}
\end{figure*}

\subsection{Predicted vs. Perceived Touristic Profile}
\label{sec:user-profile}

Besides the binary feedback of whether or not a factor of the predicted touristic profile fits well (i.e., \textit{Q06}), we also provided the possibility to adjust the respective profile via seven sliders, each for one factor of the Seven-Factor Model (i.e., \textit{Q07}).
We consider the resulting profile after \textit{Q07} as the user's perceived touristic profile.
About 90\% of the participants adjusted one or more factors of the predicted touristic profile, but on average three factors.
Similar observations are made, if the initial sample is split with respect to both aggregation strategies and analysed separately.
No statistically significant difference was observed with respect to both strategies and the number of taken adjustments.

We conducted statistical significance tests in order to capture if the Seven-Factor scores of the predicted touristic profiles differ significantly from the Seven-Factor scores of the perceived touristic profiles.
In particular, based on the outcome of the Shapiro Wilk normality test we either used Wilcoxon signed-rank test or paired Student's t-test.
Figure~\ref{fig:PredictedVsPerceivedDist} summarizes the outcome of this comparison.

Overall (N=81), the distribution of scores between predicted touristic profile and perceived touristic profiles were significantly different in factors: \textit{Sun \& Chill-Out (F1)} with p<0.05, \textit{Culture \& Indulgence (F4)} with p<0.05, \textit{Social \& Sports (F6)} p<0.01, and \textit{Nature \& Recreation (F7)} with p<0.01.
On average, the participants corrected those factors by plus 0.05-0.06. 

Focusing only on the responses to the \textit{AVG} aggregation strategy (N=41), the distributions of Seven-Factor scores of the predicted touristic profiles compared to the distributions of the perceived touristic profiles were significantly different in factors: \textit{Culture \& Indulgence (F4)} with p<0.05, \textit{Social \& Sports (F6)} p<0.01, and \textit{Nature \& Recreation (F7)} with p<0.05.
On average, the participants corrected those factors by plus 0.05-0.09.

On the other hand, by only considering the responses of participants, who were assigned the \textit{RWA} aggregation strategy (N=40), following factors showed significant differences in Seven-Factor scores distributions when the predicted and the perceived profiles were compared: \textit{Sun \& Chill-Out (F1)} p<0.05 and \textit{Nature \& Recreation (F7)} with p<0.05.
On average, the participants corrected those factors by plus 0.07-0.10.

Besides identifying differences in Seven-Factor scores distributions, we also calculated the mean absolute error (MAE), i.e., mean of the absolute differences, between predicted and perceived Seven-Factor scores, in order to capture the predictive performance of our models.
Table~\ref{tab:MAE} lists the resulting overall MAEs for each factor of the Seven-Factor model and for both strategies.
Overall, our approach showed promising performance with MAEs between 0.09 and 0.16 on a scale from 0 to 1.
Furthermore, the largest deviation from the perceived Seven-Factor scores was the one for factor \textit{Sun \& Chill-Out (F1)} with a MAE of 0.16.
For all other factors our approach showed similar performance.

Similar MAEs were observed if the predicted and perceived Seven-Factor scores were considered for both strategies separately.
Moreover, a comparison of MAEs between both strategies showed that \textit{AVG} results in a slight better predictive performance in factors \textit{Sun \& Chill-Out (F1)}, \textit{Knowledge \& Travel (F2)}, \textit{Independence \& History (F3)}, and \textit{Nature \& Recreation (F7)}.
On the other hand, \textit{RWA} slightly performed better in factors \textit{Culture \& Indulgence (F4)}, \textit{Social \& Sports (F5)}, and \textit{Action \& Fun (F6)}.
But, the differences in performance were overall not significant.

\begin{table}[h]
  \caption{Mean Absolute Error (MAE)} 
  \label{tab:MAE}
    \begin{tabular}{lrrrrrrr}
    \toprule
    {} & F1 & F2 & F3 & F4 & F5 & F6 & F7\\
    \midrule
    MAE-Overall & 0.16 & 0.10 & 0.09 & 0.10 & 0.10 & 0.10 & 0.9 \\
    MAE-\textit{AVG} & \textbf{0.16} & \textbf{0.08} & \textbf{0.07} & 0.11 & 0.12 & 0.12 & \textbf{0.07} \\
    MAE-\textit{RWA} & 0.17 & 0.13 & 0.10 & \textbf{0.09} & \textbf{0.08} & \textbf{0.09} & 0.10 \\
    \bottomrule
    \end{tabular}
\end{table}

Analysing mean absolute differences or distributional differences between Seven-Factor scores of predicted and perceived touristic profiles does not consider the user representation as a such, but rather the factors of the Seven-Factor model.
Therefore, we also took into account the distance between the predicted and perceived user representations (i.e., touristic profiles) into account.
We analysed how far apart both representations in Euclidean space are by using $DIST_{EUCL}$.
Furthermore, we used $DIST_{\tau}$ and $DIST_{SPEAR}$ to capture whether or not changes in Seven-Factor scores lead to changes in factor relevance (i.e., ranking).
For instance, after the user's predicted profile adjustment \textit{Sun \& Chill-Out (F1)} might score better and thus get more relevant (i.e., ranked higher) than \textit{Nature \& Recreation (F7)}.

We determined all three distances between predicted and perceived touristic profile for all participants and then averaged them (i.e., $\overline{DIST}_\tau$, $\overline{DIST}_{SPEAR}$, and $\overline{DIST}_{EUCL}$) in order to draw conclusions about our approach with respected to the distances.
The results are listed in Table~\ref{tab:DistPerceptionVsPredicted}. 
\textit{RWA} performed on average relatively better than \textit{AVG} with respect to the ranking of the factors (i.e., lower $\overline{DIST}_\tau$ and $\overline{DIST}_{SPEAR}$).
On the other hand, \textit{AVG} performed relatively better with respect to prediction accuracy (i.e., lower $DIST_{EUCL}$). However, the Mann-Whitney-U test showed that the differences are not significant. 

\begin{table}[h]
  \caption{
  Differences in predicted and perceived touristic profile with respect to changes in ranking (i.e., relevance) of the factors and point-wise difference.
  Note, $\overline{DIST}_\tau$ is the average $DIST_{\tau}$ between predicted and perceived touristic profiles; similarly  $\overline{DIST}_{SPEAR}$ is the average $DIST_{SPEAR}$, and $\overline{DIST}_{EUCL}$ the average $DIST_{EUCL}$.} 
  \label{tab:DistPerceptionVsPredicted}
  \begin{tabular}{lrrr}
    \toprule
     {} & Overall & \textit{AVG} &  \textit{RWA} \\
    \midrule
    $\overline{DIST}_\tau$ & 3.58 &  3.90 &  \textbf{3.25} \\
    $\overline{DIST}_{SPEAR}$ &  6.68 &  7.20 &  \textbf{6.15} \\
    $\overline{DIST}_{EUCL}$ & 0.44 &  \textbf{0.41} &  0.47 \\
    \bottomrule
  \end{tabular}
\end{table}

\section{Discussion}
\label{sec:Discussion}
Based on the responses to the demographic questions (i.e., \textit{Q08}-\textit{Q11}) the majority of participants were male, between 25 and 44 years old, and hold at least a bachelor's degree.
Thus, generalizing the outcomes and implications might only be possible in a limited way.
Future work will consider a more systematic approach of survey distribution and sampling, and eventually further distribution channels to increase and diversify the pool of participants. 

We are aware that by asking the users for their next hypothetical trip, where they have nothing to win or lose, might influence the users’ behaviour and thus introduce some bias.
Future work will consider user incentives. Ultimately, we aim at pre-trip and post-trip studies.
Furthermore, we plan to enrich the questionnaire  (e.g., by following \cite{pu2011user} and \cite{ekstrand2014user}) in order to obtain even more valuable information.

Following the approach in \cite{neidhardt2015picture}, we gave participants the option to select three to seven pictures.
Approximately half of them only selected three pictures.
This might have happened because of convenience or they may already have had a very focused idea about their next tourism destination and thus uploaded few most important pictures.
Difficulty in finding pictures might not be the reason, since the majority of the participants reported that it was easy to find pictures and also the timing indicates that they were relatively quick in selecting and ranking pictures.
Furthermore, only few participants took the chance to re-order their initial pictures selection by relevance.
Deciding, which of the selected pictures is more important than the others, might have been a difficult task.
Especially, in case of only three pictures, it might have felt unnecessary.
Another possible explanation, why users select only few pictures and do not re-consider the their ordering, is a lack of involvement, which can be addressed with user incentives in future.

Overall, our approach got positive feedback. Most of the participants were satisfied with the predicted touristic profile capturing their preferences.
The participants mostly disagreed with the predicted score for factor \textit{Sun \& Chill-Out} in comparison to all other factors.
Furthermore, no significant differences in performance could be shown between both aggregation strategies, i.e. \textit{AVG} and \textit{RWA}.
In the future, we will further improve our models, e.g., by training the CNNs systematically with more pictures.
Moreover, we plan to adapt other aggregation strategies, for instance, variations of ordered weighted averaging.

In addition to the binary feedback, whether or not a factor fits well, we also provided the option to directly adjust the predicted factors via slider inputs.
The vast majority of participants used this opportunity and adjusted at least one of the factors.
The participants adjusted the factor \textit{Sun \& Chill-Out} more often than the other factors of the Seven-Factor Model.
This is in line with the outcomes of the binary feedback (i.e., people disagreed the most with factor\textit{Sun \& Chill-Out}).
Similar observations were made by considering the mean absolute error (MAE) between predicted and perceived touristic profiles, where the biggest difference was observed in factor \textit{Sun \& Chill-Out}.
Thus, the participants not only reported that they disagree with \textit{Sun \& Chill-Out} more often than the other factors, but they also adjusted this factor the most in comparison to the others.
Overall, our approach has a tendency to underrate the predicted factor scores, 
i.e., the participants were usually correcting the predicted factors upwards.
However, based on the resulting MAEs in each factor our approach showed promising performance. 

Finally, we analysed the differences in predicted relevancy (i.e., ranking) and perceived relevancy of the factors of the Seven-Factor representation (i.e., touristic profile).
Therefore, we captured to what extend the ranking of the factors (based on the predicted Seven-Factor scores) did change after the user's adjustment.
Our results indicates that the predicted ranking is relatively close to the perceived ranking.
However, we did not consider the relative position of the rankings.
Discrepancies in top ranked (i.e., highly relevant) factors might have a  higher impact than in low ranked factors.

\section{Conclusions}
\label{sec:Conclusion}
In this paper we addressed the difficulty of travelers of explicitly expressing their preferences and needs.
We followed the idiom ``a picture is worth a thousand words'' and used pictures as a tool for communication and as a way to implicitly elicit the travelers' preferences in order to overcome communication barriers.
We designed and deployed an online user study in order to evaluate a previously introduced concept \cite{sertkan19pictures, sertkan2020pictures}, in which preferences and needs of travelers are determined based on a selection of pictures they provide.
We extended the concept by also considering the order of the pictures in the user's selection.
We asked the participants, with their next hypothetical trip in their mind, to upload three to seven pictures and rank them based on relevancy.
Based on the participants' picture selection we determined their touristic profile as a Seven-Factor representation.
We randomly and with equal chance also considered the order of the pictures in the participants' selection.
Finally, we let the participants adjust the presented profile and asked further questions.

Our user study showed promising results, as the majority of participants (65\%) were overall satisfied with their predicted touristic profile and only few (18\%) disagreed with the outcome.
Considering the order of the pictures in the participants' selection did not significantly improve the performance of our models.
The participants mostly disagreed with the factor \textit{Sun \& Chill-Out} (37\%) in comparison with the other factors (10-22\%).
Finally, we also showed that the predicted touristic profile is close to the perceived one with MAEs of factors between 0.09 and 0.16 on a scale from 0 to 1.

In future work, we will improve our approach by i) boosting up the training set; ii) considering other aggregation strategies; iii)~combining the latent feature vectors of the images with low level features (e.g., colourfulness, brightness, etc.).
Furthermore, we will focus on generalizability  by even more systematically distributing the study. Finally, we aim to compare our work to different other user modelling approaches.

\balance
\bibliographystyle{ACM-Reference-Format}
\bibliography{bibliography}

\end{document}